\documentclass[fleqn,usenatbib]{mnras}

\usepackage{newtxtext,newtxmath}
\usepackage[T1]{fontenc}

\DeclareRobustCommand{\VAN}[3]{#2}
\let\VANthebibliography\thebibliography
\def\thebibliography{\DeclareRobustCommand{\VAN}[3]{##3}\VANthebibliography}

\usepackage{graphicx}	% Including figure files
\usepackage{amsmath}	% Advanced maths commands
\usepackage{xcolor}
\usepackage{float}
\usepackage{upgreek}
\usepackage{subfig}
\usepackage{bm}
\usepackage{pifont}
\usepackage{orcidlink}
\usepackage{multirow}
\newcommand{\tick}{\ding{51}}
\newcommand{\notick}{\ding{55}}
\usepackage[normalem]{ulem}
\usepackage{cancel}
\usepackage{array}
\usepackage{dcolumn}
\usepackage{caption}
\usepackage{footnote}
\usepackage{natbib}

\title[Periodicity search in the timing of the 25 MSPs from the EPTA DR2]{Periodicity search in the timing of the 25 millisecond pulsars from the second data release of the European Pulsar Timing Array}

\author[I. C. Ni\c{t}u et al.]{\parbox{\textwidth}{Iuliana C. Ni\c{t}u\orcidlink{0000-0003-3611-3464},$^{\!1}$\!\thanks{E-mail: iuliana-camelia.nitu@manchester.ac.uk}
\,Michael~J.~Keith\orcidlink{0000-0001-5567-5492},$^{\!1}$\!\thanks{E-mail: michael.keith@manchester.ac.uk}
\,David~J.~Champion\orcidlink{0000-0003-1361-7723},$^{\!2}$
\,{Isma\"{e}l~Cognard\orcidlink{0000-0002-1775-9692},$^{\!3,4}$}\\
Gregory~\mbox{Desvignes}\orcidlink{0000-0003-3922-4055},$^{\!2}$
\,Lucas~Guillemot\orcidlink{0000-0002-9049-8716},$^{\!3,4}$
\,Yanjun~Guo,$^{\!2}$
\,Huanchen~Hu\orcidlink{0000-0002-3407-8071},$^{\!2}$
\,Jiwoong~Jang\orcidlink{0000-0003-4454-0204},$^{\!2}$\\
Jedrzej~Jawor\orcidlink{0000-0003-3391-0011},$^{\!2}$
\,{Ramesh~Karuppusamy\orcidlink{0000-0002-5307-2919},$^{\!2}$}
\,Evan~F.~Keane\orcidlink{0000-0002-4553-655X},$^{\!5}$
\,Michael~Kramer,$^{\!2}$\\
Kristen~Lackeos\orcidlink{0000-0002-6554-3722},$^{\!2}$
\,Kuo~Liu\orcidlink{0000-0002-2953-7376},$^{\!6,2}$
\,Robert~A. Main,$^{\!2}$
\,{Delphine~Perrodin\orcidlink{0000-0002-1806-2483},$^{\!7}$}
\,{Nataliya~K.~Porayko,$^{\!2}$}\\
\mbox{Golam~M.~Shaifullah\orcidlink{0000-0002-8452-4834},$^{\!7,8,9}$}
and \,Gilles~Theureau\orcidlink{0000-0002-3649-276X}$^{\!3,4,10}$
}
\\ \\ \\
\parbox{\textwidth}{$^{1}$Jodrell Bank Centre for Astrophysics, Department of Physics and Astronomy, The University of Manchester, Manchester M13 9PL, UK\\
$^{2}$Max-Planck-Institut f{\"u}r Radioastronomie, Auf dem H{\"u}gel 69, 53121 Bonn, Germany\\
$^{3}$Laboratoire de Physique et Chimie de l'Environnement et de l'Espace LPC2E UMR7328, Université d'Orléans, CNRS, F-45071 Orléans, France\\
$^{4}$Observatoire Radioastronomique de Nançay,
Observatoire de Paris, Université PSL,
CNRS, Université d’Orléans,
18330 Nançay, France\\
$^{5}$School of Physics, Trinity College Dublin, College Green, Dublin 2, D02 PN40, Ireland\\
$^{6}$ Shanghai Astronomical Observatory, Chinese Academy of Sciences, 80 Nandan Road, Shanghai 200030, China\\
$^{7}$INAF --- Osservatorio Astronomico di Cagliari, via della Scienza 5, 09047 Selargius (CA), Italy\\
$^{8}$Dipartimento di Fisica ``G. Occhialini", Universit{\'a} degli Studi di Milano-Bicocca, Piazza della Scienza 3, I-20126 Milano, Italy\\
$^{9}$INFN, Sezione di Milano-Bicocca, Piazza della Scienza 3, I-20126 Milano, Italy\\
$^{10}$Laboratoire Univers et Théories, Observatoire de Paris, Université PSL, Université de Paris Cité, CNRS, F-92190 Meudon, France\\
}
}
\date{Accepted XXX. Received YYY; in original form ZZZ}

\pubyear{2023}

\begin{document}
\label{firstpage}
\pagerange{\pageref{firstpage}--\pageref{lastpage}}
\maketitle

\begin{abstract}
In this work, we investigated the presence of strictly periodic, as well as quasi-periodic signals, in the timing of the 25 millisecond pulsars from the EPTA DR2 dataset. This is especially interesting in the context of the recent hints of a gravitational wave background in these data, and the necessary further study of red-noise timing processes, which are known to behave quasi-periodically in some normal pulsars. 
We used Bayesian timing models developed through the \textsc{run\_enterprise} pipeline: a strict periodicity was modelled as the influence of a planetary companion on the pulsar, while a quasi-periodicity was represented as a Fourier-domain Gaussian process.
We found that neither model would clearly improve the timing models of the 25 millisecond pulsars in this dataset. This implies that noise and parameter estimates are unlikely to be biased by the presence of a (quasi-)periodicity in the timing data. Nevertheless, the results for PSRs J1744$-$1134 and J1012$+$5307 suggest that the standard noise models for these pulsars may not be sufficient.
We also measure upper limits for the projected masses of planetary companions around each of the 25 pulsars. The data of PSR J1909$-$3744 yielded the best mass limits, such that we constrained the 95-percentile to $\sim\! 2 \times 10^{-4}\,\mathrm{M}_{\oplus}$ (roughly the mass of the dwarf planet Ceres) for orbital periods between 5\,d–17\,yr. These are the best pulsar planet mass limits to date.

\end{abstract}


%%%%%%%%%%%%%%%%%%%%%%%%%%%%%%%%%%%%%%%%%%%%%%%%%%

%%%%%%%%%%%%%%%%% BODY OF PAPER %%%%%%%%%%%%%%%%%%
% for terminal font: \texttt{mnras\_sample.tex}

% for in-text: \citet{Fournier1901},
% in brackets: \citep[e.g.][]{vanDijk1902}.

% equation~(\ref{eq:quadratic})
% Fig.~\ref{fig:example_figure}
% Table~\ref{tab:example_table}

%\begin{figure}
%	\includegraphics[width=\columnwidth]{example}
%    \caption{}
%    \label{fig:example_figure}
%\end{figure}

%\begin{table}
%	\centering
%	\caption{}
%	\label{tab:example_table}
%	\begin{tabular}{lccr} % four columns, alignment for each
%		\hline
%		A & B & C & D\\
%		\hline
%		1 & 2 & 3 & 4\\
%		2 & 4 & 6 & 8\\
%		3 & 5 & 7 & 9\\
%		\hline
%	\end{tabular}
%\end{table}

\section{Introduction}

Pulsar timing relies on unambiguously counting every rotation of a pulsar. 
The observed periodic pulsar emission is considered to be representative of its rotation, which is also assumed to be stable within the timescale of each observing epoch \citep[e.g.][]{Liu2012}. In practice, pulsar timing involves using a physical model of the pulsar and of the propagating medium to predict the rotational phase at each measured time of arrival (ToA hereafter), and comparing this to the observed phase at some carefully defined fiducial point in the pulsar's rotation. The difference between the observed and the modelled phase gives the \textit{timing residual}, often referred to simply as the \textit{residual}. If the model is complete, the residuals should be characterised by white noise with zero mean. 
However, excess noise is observed in many pulsar residuals in the form of long-term, time-correlated variations \citep{Cordes1985,Hobbs2010,Parthasarathy2019}, indicating that there may be additional stochastic effects that need to be included in the timing model. These can be achromatic (independent of observing frequency) spin red noise, which has been found to sometimes look quasi-periodic in slow pulsars, {and shows correlations with the observed pulse shapes in the radio emission}~\citep{Lyne2010}. Moreover, stochastic variations in the interstellar medium may cause chromatic red noise in the timing residuals. 
%Further, these variations may also indicate that the underlying assumptions about pulsar emission are incorrect.

Nevertheless, the high stability of pulses, particularly from the recycled millisecond pulsars \citep[MSPs; e.g.][]{Backer1982,Verbiest2009}, makes pulsar timing an invaluable tool for studying a range of astrophysical and cosmological phenomena, such as the structure of the interstellar medium or the solar wind, nuclear matter in exotic conditions, tests of General Relativity \citep[e.g.][]{Manchester2017} {and searches for dark matter and gravitational waves (GWs) from the early universe} \citep[e.g.][]{EPTA4}. Pulsar Timing Arrays (PTAs) represent a network of precisely and frequently timed pulsars, distributed across the Galaxy, with the purpose of detecting GWs in correlated pulsar timing signals. PTAs are expected to be sensitive to space-time distortions caused by GWs of {nanohertz frequencies}~\citep{Sazhin1978,Detweiler1979}. In this frequency range, the main contribution is believed to come from a stochastic GW background (GWB) created by the incoherent superposition of signals from in-spiralling supermassive black-hole binaries \citep[SMBHBs;][]{Rajagopal1995}. In addition, continuous GWs from strong individual SMBHB sources may also be seen in this regime~\citep{Estabrook1975}, as well as other, more exotic theoretically predicted GW sources such as cosmic strings or a relic GW background~\citep{Kibble1976, Grishchuk2005}. 
Several collaborations have been developed with the specific purpose of detecting these GWs, including the European Pulsar Timing Array \citep[EPTA;][]{Janssen2008, Kramer2013}, the Parkes Pulsar Timing Array \citep[PPTA;][]{Hobbs2010, Manchester2013}, the North American Nanohertz Observatory for Gravitational Waves \citep[NANOGrav;][]{Demorest2013, Arzoumanian2015}, the Indian Pulsar Timing Array \citep[InPTA;][]{Joshi2018}, the Chinese Pulsar Timing Array \citep[CPTA;][]{Lee2016}, and the MeerTime Pulsar Timing Array \citep{Spiewak2022}. These groups also work together as part of the International Pulsar Timing Array \citep[IPTA;][]{Hobbs2010, Manchester2013, Verbiest2016}.

Recently, results on the latest search for a GWB signature were published concurrently by the EPTA+InPTA \citep{EPTA2023III}, PPTA \citep{Reardon2023}, NANOGrav \citep{Agazie2023}, and CPTA \citep{Xu2023}. These reported an emerging evidence of a stochastic GWB in their datasets, ranged between $2\text{--}5\sigma$ in significance, depending on the PTA. A comparison and initial effort to combine these analyses can be found in \citet{IPTA2023}. However, these analyses do not yet meet the requirements for a clear GWB detection. Furthermore, there are still effects in the data of all PTAs that are not yet fully understood, such as how to best model red noise variations in the timing residuals \citep[e.g.][]{EPTA2023II}. 

In this work, we study the individual-pulsar timing models of the EPTA Second Data Release (DR2), as used in the GWB search \citep{EPTA2023I,EPTA2023II,EPTA2023III}. To perform the analyses, we use the Bayesian toolkit \textsc{run\_enterprise} \citep{run_enterprise} based on the pulsar timing frameworks \textsc{enterprise} \citep{Ellis2019} and \textsc{tempo2} \citep{Edwards2006}. Specifically, we investigate the consequences on the timing of each of the 25 MSPs of adding a strictly periodic, planet-like component, as well as a Fourier-domain quasi-periodic (QP) component to the Bayesian model fitting. The latter is particularly interesting since, as shown in \citet{Keith2023} for slow pulsars, unmodelled QP behaviours present in pulsar timing data can affect the robustness of some parameter and noise estimates. 

Furthermore, the sensitivity of this dataset provides excellent constraints on the masses of any putative pulsar planetary companions.
Although there have been several systematic searches in the timing of pulsars \citep{Thorsett1992,Kerr2015,Behrens2020,Nitu2022}, only six pulsars have been confirmed to host planetary-mass companions. The most famous, PSR B1257$+$12, has three low-mass companions, of $0.020(2)\,\mathrm{M}_{\oplus}$, $4.3(2)\,\mathrm{M}_{\oplus}$ and $3.9(2)\,\mathrm{M}_{\oplus}$, with orbital periods of $25.262(3)\,\mathrm{d}$, $66.5419(1)\,\mathrm{d}$ and $98.2114(2)\,\mathrm{d}$, respectively \citep{Wolszczan1992,Wolszczan1994,Konacki2003}. PSR B1620$-$26 is in a triple system located in the globular cluster M4, also containing a white dwarf and a $\sim\!\! 2.5$ Jupiter-mass planet ($\sim\!\!800\,\mathrm{M}_{\oplus}$) of orbital period $\sim\!\! 36\,500\,\mathrm{d}$~\citep{Thorsett99}. PSRs J1719$-$1438 \citep{Bailes2011}, J0636$+$5128 \citep{Stovall2014}, J1311$-$3430 \citep{Romani2012, Pletsch2012}, and J2322$-$2650 \citep{Spiewak2018} each have one `diamond planet'\footnote{The so-called `diamond planets' are ultra-low mass carbon white dwarfs, believed to be the remains of a disrupted stellar companion \citep[e.g.][]{Bailes1991}.} companion of mass between $\sim\!\!1 \text{--}10$ Jupiter-mass ($\sim\!\!300 \text{--} 3200\,\mathrm{M}_{\oplus}$) and orbital period $<\!1\,\mathrm{d}$. The rarity of pulsar planetary companions is likely a consequence of the extreme conditions in which pulsars form. There is currently no clear mechanism(s) to creating these systems --- for an overview of proposed scenarios and formation paths, see e.g. \citet{Podsiadlowski1993}, \citet{Phillips1994} and the Introduction of \citet{Nitu2022}. 
Recently, \citet{Nitu2022} performed the largest scale search for pulsar planetary companions using the timing datasets of 800 pulsars observed at the Jodrell Bank Observatory (JBO). They estimated that at most 0.5\% of all pulsars are expected to host Earth-mass planets, concluding that planets around pulsars must be extremely rare. Separately, \citet{Behrens2020} conducted a search on the NANOGrav 11-yr dataset, and found no evidence of planetary companions around any of the 45 MSPs. They also estimated the planet-mass sensitivity of the NANOGrav dataset, as a function of orbital period. 
In this analysis, we follow the analysis method in \citet{Nitu2022} to search for and assess the sensitivity of the EPTA DR2 dataset to the influence of planetary companions. We then also directly compare our results with those of \citet{Behrens2020}.

This paper is structured as follows. In Section~\ref{sec: dataset}, the properties of the EPTA DR2 dataset are summarised. Section~\ref{sec: coretiming} describes the main properties of the timing model used throughout this work. In Section~\ref{sec: planetfits} we outline the planet-fitting method, and present and discuss the corresponding results. Section~\ref{sec: QPfit} presents the setup and discusses the results of the QP fitting. In Section~\ref{sec: conclusions} we summarise our conclusions.

\section{Dataset} \label{sec: dataset}

The EPTA DR2 is one of the current state-of-the-art pulsar timing datasets, having been recently used in the search for a GWB signature \citep{EPTA2023III}. It contains high-precision pulsar timing data from 25 MSPs, collected over 25\,years, with five large telescopes in Europe: the 100-m Effelsberg Telescope (in Germany); Jodrell Bank Observatory's 76-m Lovell Telescope (in the United Kingdom); Nan\c{c}ay Radio Observatory's large Radio Telescope (NRT; in France); the Astronomical Observatory of Cagliari's 64-m Sardinia Radio Telescope (SRT; in Italy); the Westerbork Synthesis Radio Telescope (WSRT; in the Netherlands). 
Once a month, these telescopes also functioned collectively, as the Large European Array for Pulsars (LEAP), which is equivalent to a 194-m sixth interferometric telescope in the EPTA \citep{Bassa2016}. 
Most of the observations part of the EPTA DR2 are at \mbox{`L-band'} frequencies ($1\text{--}2\,\mathrm{GHz}$) and above, with bandwidths of up to $512\,\mathrm{MHz}$. A limited number of observations are centred at lower frequencies of $350\,\mathrm{MHz}$. For a detailed description of the properties of the EPTA DR2 dataset, see \citet{Chen2021}, and \citet{EPTA2023I}.

\color{black}
\section{Core timing model} \label{sec: coretiming}

Throughout this work, we use the \textsc{run\_enterprise} pipeline \citep{run_enterprise} on the timing data of each pulsar. In this Bayesian framework, all the components of the desired model are fit for simultaneously. In the following sections, we discuss, separately, two additional model components which have not been considered in the standard GWB analyses: a simple periodicity (planet) component (Section~\ref{sec: planetfits}); and a QP Gaussian process component (Section~\ref{sec: QPfit}). In both cases, however, the analysis includes the same typical `core' (base) timing model, which is summarised in this section.

The core timing model used here follows the same principles as in the individual pulsar analysis of the EPTA DR2 dataset \citep{EPTA2023II}. The deterministic components of the pulsar model (i.e. the effects of spin frequency and its derivatives, pulsar position, known binary companions, etc.) are fit for simultaneously with the noise components. The parameters describing the former are generally marginalised over for efficiency. {The noise model consists of a white noise component as well as between one and three Fourier-domain red noise terms.
The white noise is modelled using the EFAC and EQUAD parameters which scale and add in quadrature to the estimated measurement error (see e.g. \citealp{Verbiest2016} for more details); theseccount for unmodelled instrumental errors and intrinsic pulse jitter  in the arrival times~\citep{Liu2012,Parthasarathy2021}. The EPTA DR2 dataset is composed of primarily narrow-band observations and does not use the ECORR parameter \citep{hv14}.
The choice of red noise model makes use of the model selection process in \citet{EPTA2023II}. The red noise components are chosen from a combination of:} (i) an achromatic red noise term; (ii) a chromatic Dispersion Measure (DM) term inducing a timing delay with an $f_\mathrm{obs}^{-2}$ dependence on the observing frequency $f_\mathrm{obs}$; and (iii) a chromatic scattering variation (SV) term, with an $f_\mathrm{obs}^{-4}$ dependence. These are all implemented following the method in \citet{Lentati2014} and using power-law priors for the corresponding power spectral density (PSD), i.e.
\begin{equation}
    \label{eq: PL}
    P(f) = \frac{A^2}{K} \left(\dfrac{f}{1\,\mathrm{yr}^{-1}}\right)^{-\gamma},
\end{equation}
{where $f$ is the Fourier frequency, $A$ and $\gamma$ are the power-law amplitude and slope (index), respectively, and $K$ is a scale factor for each process.
In total there are six hyperparameters: the slopes for each process, $\gamma_\mathrm{red}$, $\gamma_\mathrm{DM}$, $\gamma_\mathrm{SV}$; and the corresponding log-amplitudes, $\log_{10}\!A_\mathrm{red}$, $\log_{10}\!A_\mathrm{DM}$, $\log_{10}\!A_\mathrm{SV}$. For the achromatic and SV processes, $K=12\pi^2$, and for the DM process $K=k_\mathrm{DM}^2$, with the DM constant $k_\mathrm{DM}=2.41\times10^{-4} \mathrm{cm^{-3}pc\,MHz^2s^{-1}}$. For more details see \citet{EPTA2023II}.
The Fourier basis has equally spaced frequencies $f_n = n/T_\mathrm{span}$, with $n \!\in\! \{1, 2, \dots, N_\mathrm{c}\}$, $T_\mathrm{span}$ the total time span of the data, and $N_\mathrm{c}$ the number of Fourier components. In this study, we set $N_\mathrm{c}$ for each of the achromatic, DM and scattering red noise to the `optimal' values for this dataset as found by \citet{EPTA2023II}.}

\section{Fitting for a planet influence} \label{sec: planetfits}

\subsection{Setup}
\color{black}
We use the Bayesian planet fitting model implemented in \textsc{run\_enterprise}, which is introduced and described in \citet{Nitu2022}. In short, this is parameterised by the (fitted) orbital parameters: projected mass\footnote{Note that whenever we talk about a planetary-companion `mass' in this work, we mean the projected mass, as it is not possible to disentangle the inclination dependence in this type of analysis without additional information.} of the planet, $m\sin{i}$ (where $m$ is the planetary mass and $i$ the inclination of the orbit); orbital period, $P_\mathrm{b}$; eccentricity, $e$; argument of periapsis, $\omega$; and the phase $\phi$ of the planet on the orbit with respect to the periastron crossing, defined at a reference time $t_\mathrm{ref} = 55000\,\mathrm{MJD}$. These parameters all contribute to the R\o{}mer delay, which quantifies the planetary influence on the pulsar signal for a ToA $t$. This can be expressed as \citep[e.g.][]{Blandford1976}
\begin{equation}
    \Delta_\mathrm{R} (t)  = \mathcal{C}\, P_{\mathrm{b}}^{2/3}\,m \sin{i} \, [(\cos{E(t)}-e)\sin{\omega} + \sin{E(t)}\sqrt{1-e^2}\cos{\omega}],
    \label{eq: roemer}
\end{equation}
where $\mathcal{C}$ is a constant of proportionality. We set the mass of each pulsar to be fixed at a characteristic value of $M_\mathrm{PSR}\!=\! 1.4\,\mathrm{M}_{\odot}$ \citep[e.g.][]{Lattimer2012}, and to be much larger than the mass of the planet, such that $\mathcal{C} \simeq 23.4 \,\mathrm{d}^{-2/3}\mathrm{M}_{\oplus}^{-1}\upmu\mathrm{s}$ with the R\o{}mer delay generally expressed in microseconds, $P_\mathrm{b}$ in days, and $m$ in Earth masses ($\mathrm{M}_{\oplus}$).
The eccentric anomaly $E(t)$ in Eq.~\ref{eq: roemer} is related to the true ($\mathcal{A}_\mathrm{T}(t)$) and mean ($\mathcal{M}(t)$) anomalies by
\begin{equation}
    \cos{E(t)} = \frac{e+\cos{\mathcal{A}_\mathrm{T}(t)}}{1+e\cos{\mathcal{A}_\mathrm{T}(t)}}
    \label{eq: cosE}
\end{equation}
and
\begin{equation}
    \mathcal{M}(t) \equiv \frac{2\pi}{P_{\mathrm{b}}} \, (t-t_0) = E(t) - e\sin{E(t)},
    \label{eq: meanan}
\end{equation}
where $t_0$ is the time of closest periastron approach. To find an equation for $E(t)$ that is only a function of $t$ and the fitted orbital parameters (in this case $e$ and $\phi$), we start from the definition of $\phi$, i.e. $\mathcal{A}_\mathrm{T}(t_\mathrm{ref}) = 2\pi\phi$. It follows that $E(t_\mathrm{ref})$ can be obtained from $\phi$, $e$, and Eq.~\ref{eq: cosE} at $t\equiv t_\mathrm{ref}$. Further, from Eq.~\ref{eq: meanan} at $t \equiv t_\mathrm{ref}$, we can compute
\begin{equation}
    t_0 = t_\mathrm{ref} - \frac{P_{\mathrm{b}}}{2\pi} [E(t_\mathrm{ref}) - e\sin{E(t_\mathrm{ref})}].
    \label{eq: t0}
\end{equation}
Finally, $E(t)$ can consequently be estimated for any $t$ by solving the non-linear expression in Eq.~\ref{eq: meanan}.

The Bayesian priors of the five orbital parameters determining the planet influence ($m\sin{i}$, $P_\mathrm{b}$, $e$, $\omega$, and $\phi$) are set up as follows. Uniform priors are used for $\omega\!\in\![0, 2\pi)$ and $\phi\!\in\![0, 1)$. Log-uniform priors are used to explore the parameter space for $e\!\in\![0,0.9]$, and the projected mass $m\sin{i}\!\in\![10^{-5}, 10^{-1}]\,\mathrm{M}_{\oplus}$. Note that we search for lower planetary masses than in the \citet{Nitu2022} analysis since we expect the EPTA DR2 dataset to be more sensitive to such influences than the Jodrell Bank Observatory dataset by itself. We split the parameter space of $P_\mathrm{b}$ into 11 period bins, to thoroughly explore the large prior range; the bounds of these are $\{5.0, 10.1, 21.3, 42.5, 85, 170, 340, 390, 780, 1560, 3120, 6240\}$\,d. Ten of these bins are log-uniformly spaced, while one narrower bin (340--390\,d) is considered around $P_\mathrm{b} = 1\,\mathrm{yr}$ to account for the loss of sensitivity due to also fitting for the pulsar position and parallax (which have a 1-yr periodicity) in the same analysis.

The EPTA DR2 timing data for each of the 25 MSPs are initially processed through the pipeline \textsc{run\_enterprise}, including the `core' and planet model, using the \textsc{python} Markov Chain Monte Carlo (MCMC) sampler \textsc{emcee} \citep{Foreman-Mackey2013, Foreman-Mackey2019}. From the posteriors, planetary mass limits are then estimated at the 95\% threshold for each period bin in each pulsar. Potential planet candidates are selected based on a 3-$\sigma$ `flag' in the linear mass posterior, i.e. when the mean of the distribution is more than three times the standard deviation. For more details on these processes, see \citet{Nitu2022}.
Moreover, for each of these `flagged' pulsars, a further nested-sampling run is performed, using the \textsc{python} sampler \textsc{dynesty} \citep{Speagle2020}. This nested sampling method yields the Bayesian evidence in favour of including the planet model, as well as providing a consistency check for the parameter values obtained from the MCMC run. Further investigations are then made into the credibility of the potential planet detection, by inspecting the residuals, and the PSD plots in this context.

\subsection{Results \& Discussion}
\subsubsection{The 3-$\sigma$ flagged MSPs} \label{sec: 25PSRs_flagged}
\color{black}

For 4 of the 25 MSPs in this study, the planet-mass posterior distribution in one period bin has the property that the mean is more than three standard deviations away from zero. We call these the `flagged' MSPs, or those showing `3-$\sigma$ detections'. 
Three of these MSPs are known to be in stellar binary systems, with orbital periods: 0.26\,d for PSR J0751$+$1807; 0.60\,d for PSR J1012$+$5307; and 10.91\,d for PSR J1918$-$0642. The former two are below the minimum orbital period that we search for in this analysis (5\,d) and are therefore unlikely to affect this substantially. The latter may reduce the sensitivity of the lowest period bins we use. The fourth `flagged' MSP, PSR J1744$-$1134, is a solitary pulsar.

Table~\ref{tab: flaggedPSRs} shows the projected mass and orbital period of the potential planet companion for each pulsar, as estimated from the posteriors of the separate MCMC (\textsc{emcee}) and nested sampling (\textsc{dynesty}) runs, respectively. The eccentricity is not quoted as the posterior distributions generally recover the prior for all pulsars. 
% The posterior distributions of the orbital parameters for these 10 MSPs can be seen in Appendix~\ref{app: cornerplots} in the shape of corner plots, for both the MCMC and nested-sampling methods.
\begin{table*}
	\centering
	\caption{A summary of the main properties of the periodicity analysis on the 4 MSPs showing 3-$\sigma$ detections in the planet mass posterior. The second column shows the natural-logarithm Bayes factor (evidence) in favour of including the planet model to the `core' timing model. The mean and standard deviation values are given for the planetary projected mass, $m\sin{i}$, and orbital period, $P_\mathrm{b}$. The uncertainties are given in standard parenthetical notation, i.e. representing one standard deviation in the last digit. The inverse of the maximum-likelihood fundamental frequency of the QP analysis ($f_\mathrm{qp}$) is given for comparison (`maxL'); the standard deviation (`stdev') of the $1/f_\mathrm{qp}$ posterior is also included, but should only be used to give an idea of the spread of values; see also Fig.~\ref{fig: QP_cornerplots} for the shape of these posteriors. The total time of the observations, $T_\mathrm{span}$, is included for a comparison with the quoted periodicities. Finally, the choice of noise models (achromatic `Red' noise, `DM' Noise, and scattering variation `SV') used in the analysis of each pulsar is shown, as determined by \citet{EPTA2023II}.}
	\label{tab: flaggedPSRs}
	\begin{tabular}{cccccccccc} % four columns, alignment for each
		\hline
		\multirow{2}{*}{PSR} & \multirow{2}{*}{$\ln\!\mathcal{B}$} & \multicolumn{1}{c}{\;\;$m\sin{i} \,[\!10^{-4}\, \mathrm{M}_{\oplus}\!]$} & \multirow{1}{*}{$P_{\mathrm{b}} [\mathrm{d}]$} & \multicolumn{2}{c}{$1/f_\mathrm{qp}[\mathrm{d}]$} & \multicolumn{1}{c}{$T_\mathrm{span}$} & \multicolumn{3}{c}{Noise models}\\
		&  & \multicolumn{1}{c}{\;\;\;\,\textsc{emcee} | \textsc{dynesty}} & \multicolumn{1}{c}{\;\;\;\,\textsc{emcee} | \textsc{dynesty}} & maxL & stdev & \multicolumn{1}{c}{$[\mathrm{d}]$} & Red & DM & SV\\
		\hline
        J0751$+$1807 & \;\,3.1(8) & 1.5(6) | 1.6(6) & 2490(230) | 2460(210) & 1940 & 630 & 8835 & \notick & \tick & \notick \\
        J1012$+$5307 & \;\,0.5(9) & 1.6(7) | 1.0(9)& \,\,\,760(30) | 730(130)& \,\,\,770 & 560 & 8648 & \tick & \tick & \notick \\
        % \:\:\:\:\:\,J1600$-$3053\,(a) & \;\,\textbf{5.5(7)} & 0.8(3) | 0.9(3) & 1070(50) | 1080(50) & \multirow{2}{*}{1040} & \multirow{2}{*}{350} & \multirow{2}{*}{5232} \\
        % \:\:\:\:\:\,J1600$-$3053\,(b) & \textbf{16.7(7)} & 0.8(3) | 1.0(2)& 2350(210) | 2600(250)&  &  &  \\
        J1744$-$1134 & \;\,0.4(8) & 0.6(2) | 0.4(3) & \,\,\,1550(70) | 1480(200) & 1460 & 500 & 8770 & \tick & \tick & \notick\\
        J1918$-$0642 & \;\,1.7(7) & 1.5(1.4) | 2.4(1.3) & 3000(570) | 2940(330)& 720 & 480 & 7199 & \notick & \tick & \notick\\
    	\hline
	\end{tabular}
\end{table*}
The estimated log-Bayes evidence in favour of including a planet influence in the timing model for each pulsar is also shown in Table~\ref{tab: flaggedPSRs}. As all log-Bayes factors are within four standard deviations of 0, these values suggest that there is no support for the planet model from the Bayes factors.

We note that the masses shown in Table~\ref{tab: flaggedPSRs} are roughly two orders of magnitude below even the smallest known pulsar planet, i.e. the $0.02\text{-}\mathrm{M}_{\oplus}$ companion of PSR B1257$+$12 \citep{Wolszczan1992,Wolszczan1994,Konacki2003}. On the other hand, the orbital periods are two orders of magnitude \textit{above} those of the planets of PSR B1257$+$12 ($<\!100\,\mathrm{d}$), such that the inferred pulsar--planet distances would be similarly larger in these systems. This is somewhat expected, as our dataset is highly sensitive to these kinds of behaviours. Interestingly, the $800\text{-}\mathrm{M}_{\oplus}$ planet in the triple-system B1620$-$26 has an orbital period roughly an order of magnitude above those of our `flagged' pulsars, three of which are also known to be in binary systems \citep{Thorsett99}.

\begin{figure} 
    \centering
        \includegraphics[width=\linewidth]{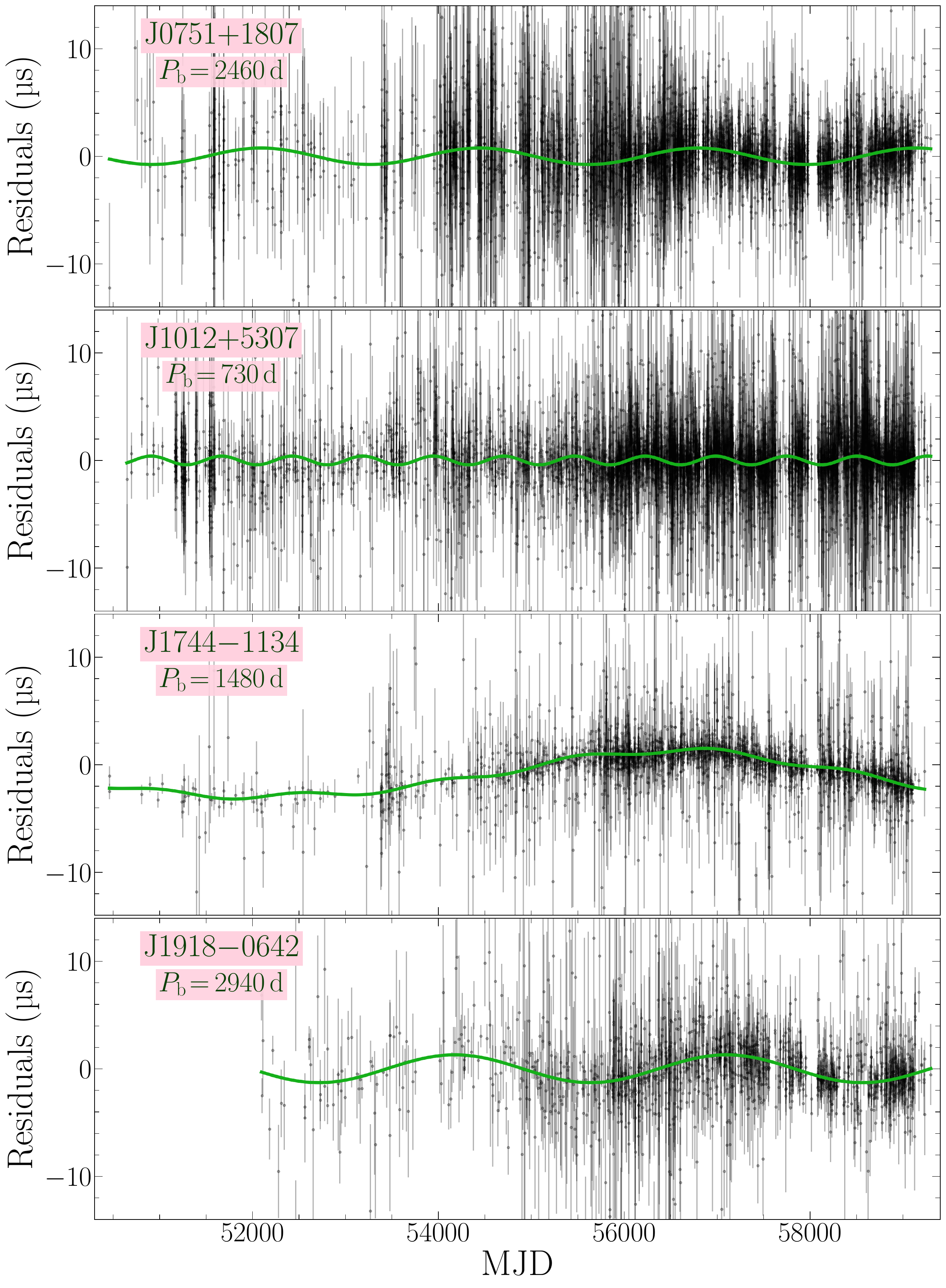}
    \caption{The timing residuals, without removing the possible planet influence, of the 4 MSPs that showed a 3-$\sigma$ flag in the planet fitting (black data points), and the corresponding maximum-likelihood R\o{}mer delay (green curve), for comparison. Note that for PSR J1744$-$1134 only, the green curve also includes the lowest two Fourier-frequency components of the achromatic red noise to aid the direct comparison. See text for more details on how these residuals have been computed.}
    \label{fig: flagres}
\end{figure}

\begin{figure} 
    \centering
    \includegraphics[width=\linewidth]{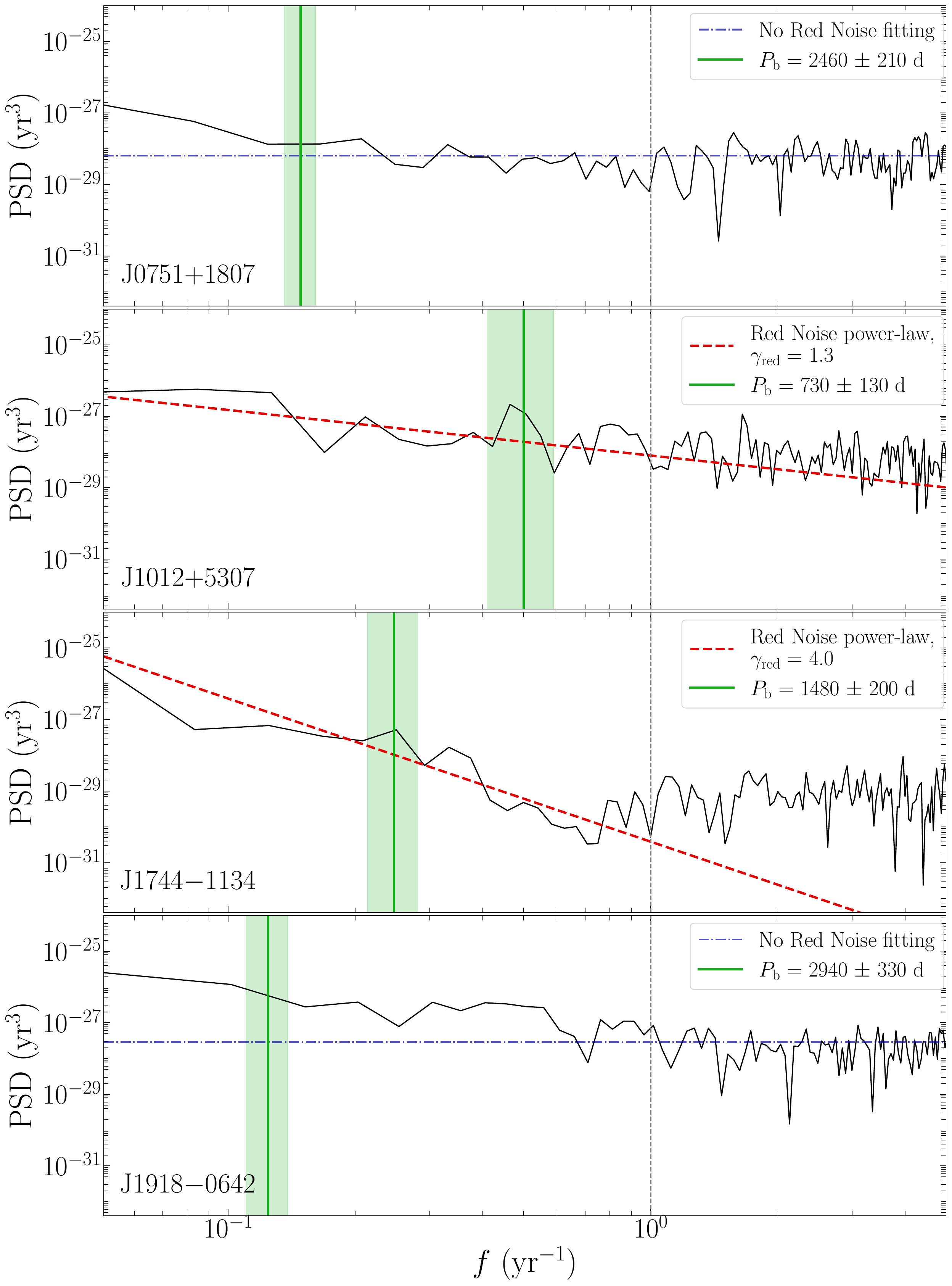}
    \caption{The estimated PSD as a function of Fourier frequency (black line); the best-fit power-law red noise (red dotted line), {or white noise level (blue dot-dashed line)} is shown, as well as the orbital period detected (vertical green line), all for each of the 4 `flagged' MSPs. The grey dashed vertical line represents the 1/yr frequency.}
    \label{fig: flagPSD}
\end{figure}

To further understand these potential planet detections and their cause, we look at how the fitted planet influence compares to the residuals of each of the four pulsars (Fig.~\ref{fig: flagres}), as well as the shape of the PSD of the achromatic red noise modelled in these residuals in the absence of the planet fitting (Fig.~\ref{fig: flagPSD}). 

The choice of noise models used for each of the four MSPs is summarised in Table~\ref{tab: flaggedPSRs}, as per the modelling selected in \citet{EPTA2023II}. Correspondingly, the residuals are computed by subtracting from the observed ToAs the modelled DM time series and the best deterministic timing model, except for the planet influence. This is enough to show the left-over planet influence. Note that the modelled noise that is removed is the best-fit model to the data in the simultaneous Bayesian fit which included the planet influence. Thus the resulting residuals should only contain white noise, any achromatic red noise, and the planet influence, according to the best timing model; these residuals can therefore be compared to the analytical R\o{}mer delay as computed from the best-fit orbital parameters. Note that for PSR J1744$-$1134 --- the only one of the four to have significant achromatic red noise --- we compare the residuals obtained as above with the analytical R\o{}mer delay to which we also add the lowest two Fourier-frequency achromatic red noise components.
These comparisons are shown in Fig.~\ref{fig: flagres}.

In these plots, perhaps the most noticeable property is that the possible planet influence is generally of small amplitude, similar in value to the white noise level, and therefore largely indistinguishable from it. This is not unexpected following the mass and orbital period values quoted in Table~\ref{tab: flaggedPSRs}, as a companion of mass $\sim\!10^{-4}\,\mathrm{M}_{\oplus}$ orbiting with a period of order 1\,yr induces a maximum variation of order $0.1\,\upmu\mathrm{s}$ in the pulsar ToAs (estimated from the R\o{}mer delay as in Eq.~\ref{eq: roemer}). 
For the two pulsars of highest Bayesian evidence (J0751$+$1807 and J1918$-$0642) the detected periodicity is large compared to the total time span of the data, such that less than 4 full periodicities are seen in the analysed dataset. It is therefore not clear whether this observed variation will continue to behave as a simple periodicity, or will take the form of a red-noise like process in a longer dataset.

It is also interesting to investigate the shape of the fitted achromatic red noise in the absence of the planet fitting model, particularly in the Fourier domain. We expect any noticeable periodicity in the residuals to show as a peak in the corresponding PSD of these residuals, which may also bias a simple power-law fit. To estimate these PSD shapes from the residuals, we employ the widely-used `Cholesky method' as described in \citet{Coles2011}, and also adopted in \textsc{cholspectra}. In short, this is based on estimating the covariance matrix of the residuals assuming the power-law form of the PSD (as in Eq.~\ref{eq: PL}), then using the Cholesky decomposition on this covariance matrix to determine the transformation that `whitens' the residuals. Thus the fitting problem becomes a simple ordinary least-squares on uncorrelated data, and the PSD can be straight-forwardly estimated from the best-fit parameters. The appropriate residuals are obtained by subtracting the best-fit deterministic timing model, as well as any fitted DM and SV noise, from the pulsar signal; note that this is now in the case of a model \textit{not} including a planet influence. These PSD estimates are seen in Fig.~\ref{fig: flagPSD}, together with the best power-law model (or lack there-of), and the maximum-likelihood $P_\mathrm{b}$ from the planet-fitting runs.

Similarly to the findings from the residual plots of these pulsars, any tentative peak in the PSD at the `detection' orbital period cannot be unambiguously distinguished from a stochastic variation around the power-law prior. 
While for PSRs J0751$+$1807 and J1918$-$0642 the best noise model as per \citet{EPTA2023II} did not include an achromatic term, the estimated PSD appears to have some excess power at very low Fourier frequencies compared to a flat (`white') power. 
{For both of these pulsars, comparing the DM-only model to a combination of DM and achromatic noise favours the combined model, but with an insignificant Bayes factor of ${\ln\mathcal{B}\sim1}$.
The planet-like periodicities found for these two pulsars are similar to the time span of their data, and the evidence for the planetary model is similar to that for an achromatic power-law noise model.
Therefore we argue that the simplest conclusion is that these pulsars have a low level of achromatic noise that was not included in the base model, which can equally well be modelled by a single low-frequency sinusoid or a power-law process.
If we repeat the analysis with the addition of an achromatic power-law model, there is no evidence in favour of a planetary companion.
Given the prevalence of power-law red noise in pulsars, and the scarcity of planetary companions, it seems most plausible that this is simply unmodelled red noise.}

%However, lack of evidence in favour of either the periodic or power-law noise suggests that the apparent low-frequency noise may not really be timing noise, and rather may be related to the fitting for chromatic noise. As found by \citet{Keith2013}, a large uncertainty in the DM noise model may lead to an excess `signal' in the PSD. On the other hand, for PSRs J1012$+$5307 and J1744$-$1134, tentative peaks are seen at the detected planet-like periodicity, although many such features can be seen at various Fourier frequencies.
% It is also worth noting that in PSRs J1012$+$5307, our fitting recovers a planet orbital period that is very close to 2\,yr. This may be a coincidence, or may be due to some unmodelled effects in the pulsar timing model, such as the yearly variable proper motion. 

To summarise, given the discussed properties and mass limits obtained, there is no evidence of planetary companions in this dataset, and the planet-like periodicities found in this analysis are likely an artefact of the choice of noise models. However, the relatively large achromatic red-noise power on timescales close to the dataspan is worth investigating further in future work. Interestingly, the PSRs J1012$+$5307 and J1744$-$1134 highlighted in this analysis are also characterised as having some level of complex behaviours in the recent noise analysis of the EPTA DR2 dataset \citep{EPTA2023II}. Indeed, we notice in the PSD shapes of these pulsars that there is additional complexity with respect to a single power-law model. \citet{EPTA2023II} propose that these effects may be due to e.g. a non-stationarity of the stochastic red noise, or perhaps some unknown instrumental effects. They also suggest that the observed noise properties in these pulsars should be further studied with a different dataset such as the upcoming IPTA data combination, in an attempt to fully understand their behaviours.

\subsubsection{Companion mass limits}
\color{black}
\begin{figure} 
    \centering
    \includegraphics[width=\linewidth]{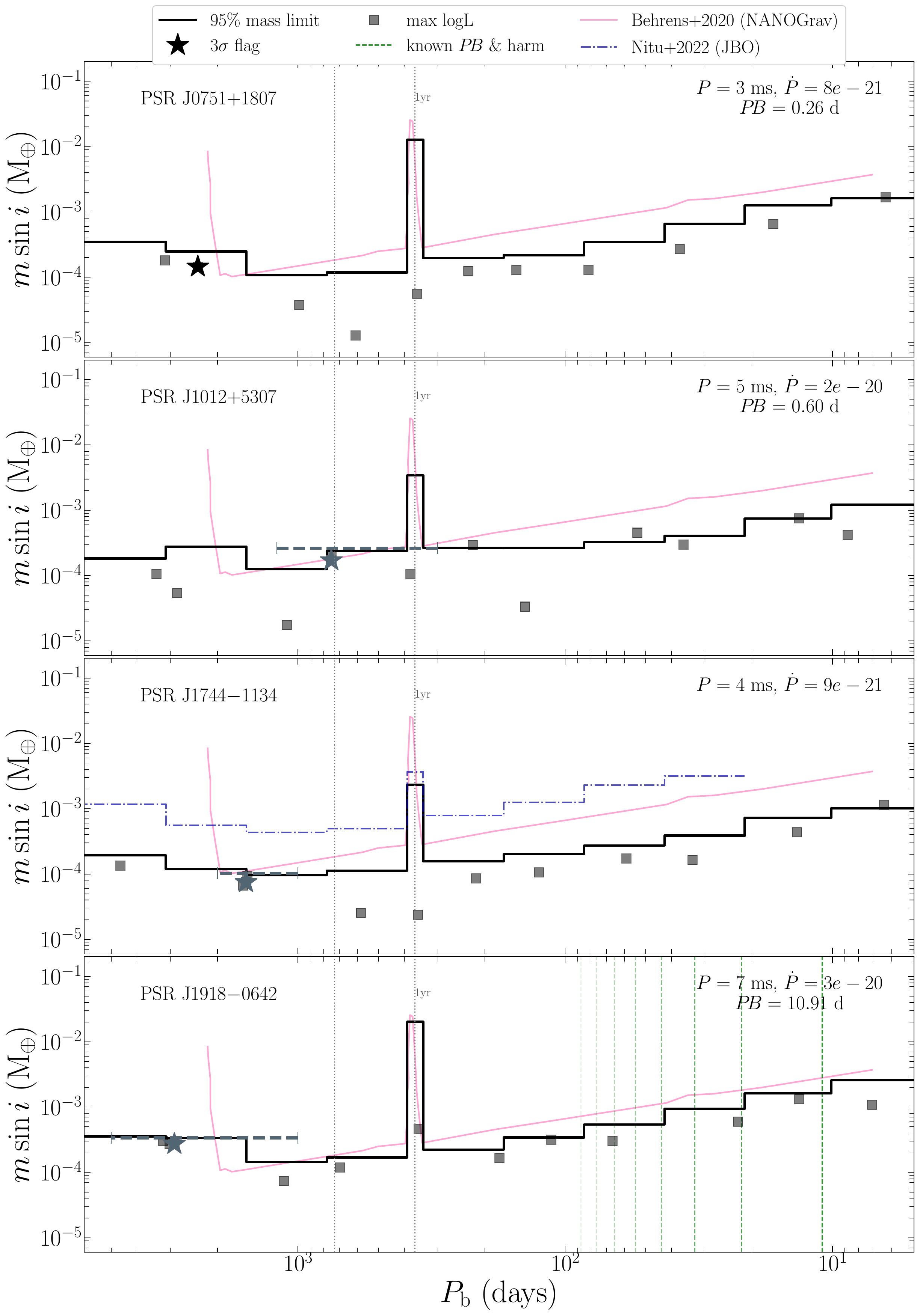}
    \caption{The 95\% planet mass limits for each orbital period bin, as estimated from the planet fitting results for the 4 `flagged' MSPs (black step line). The properties in the top right are the spin-period of the pulsar and its derivative, as well as the orbital period of the known binary companion. The black squares represent the maximum-likelihood values in each period bin, while the stars represent 3-$\sigma$ detections. The additional runs at particular period bins were informed by the initial results, and are shown in horizontal grey lines for PSRs J1012$+$5307 and J1744$-$1134. The vertical green dashed lines show the fundamental periodicity and harmonics of the known stellar companions of the respective pulsars. The pink line shows the sensitivity from the NANOGrav 11-yr dataset, as estimated by \citet{Behrens2020}. In PSR J1744$-$1134, the blue line represents the mass limits as estimated only from the JBO data, as given in \citet{Nitu2022}.}
    \label{fig: flagmasslim}
\end{figure}
\begin{figure} 
    \centering
    \includegraphics[width=\linewidth]{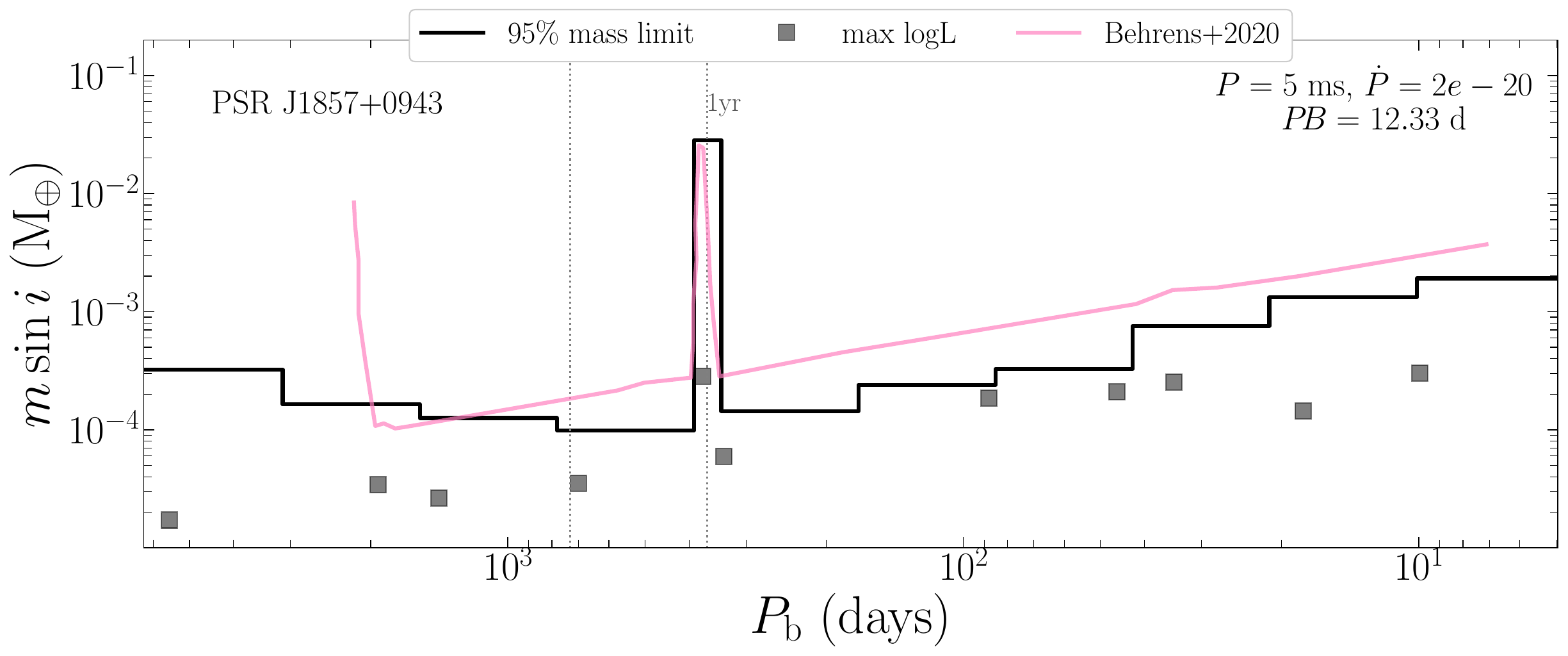}
    \caption{The 95\% planet mass limits for each orbital period bin, as estimated from the planet fitting results for the most-sensitive MSPs in this dataset, PSR J1909$-$3744. The pink line shows the sensitivity from the NANOGrav 11-yr dataset, as estimated by \citet{Behrens2020}.}
    \label{fig: J1909masslim}
\end{figure}
It is also interesting to consider the planet-mass upper limits that can be inferred from the planet search, as we expect this dataset to be highly sensitive to such influences. For each of the 11 period bins that we perform our analysis on, we can estimate the 95\% upper limit on the projected mass, simply from the posterior distribution of the mass. Note that the sampling gives the log-mass posterior, which is then converted to the linear-mass posterior for this estimate.

\begin{figure*} 
    \centering
    \includegraphics[width=0.85\linewidth]{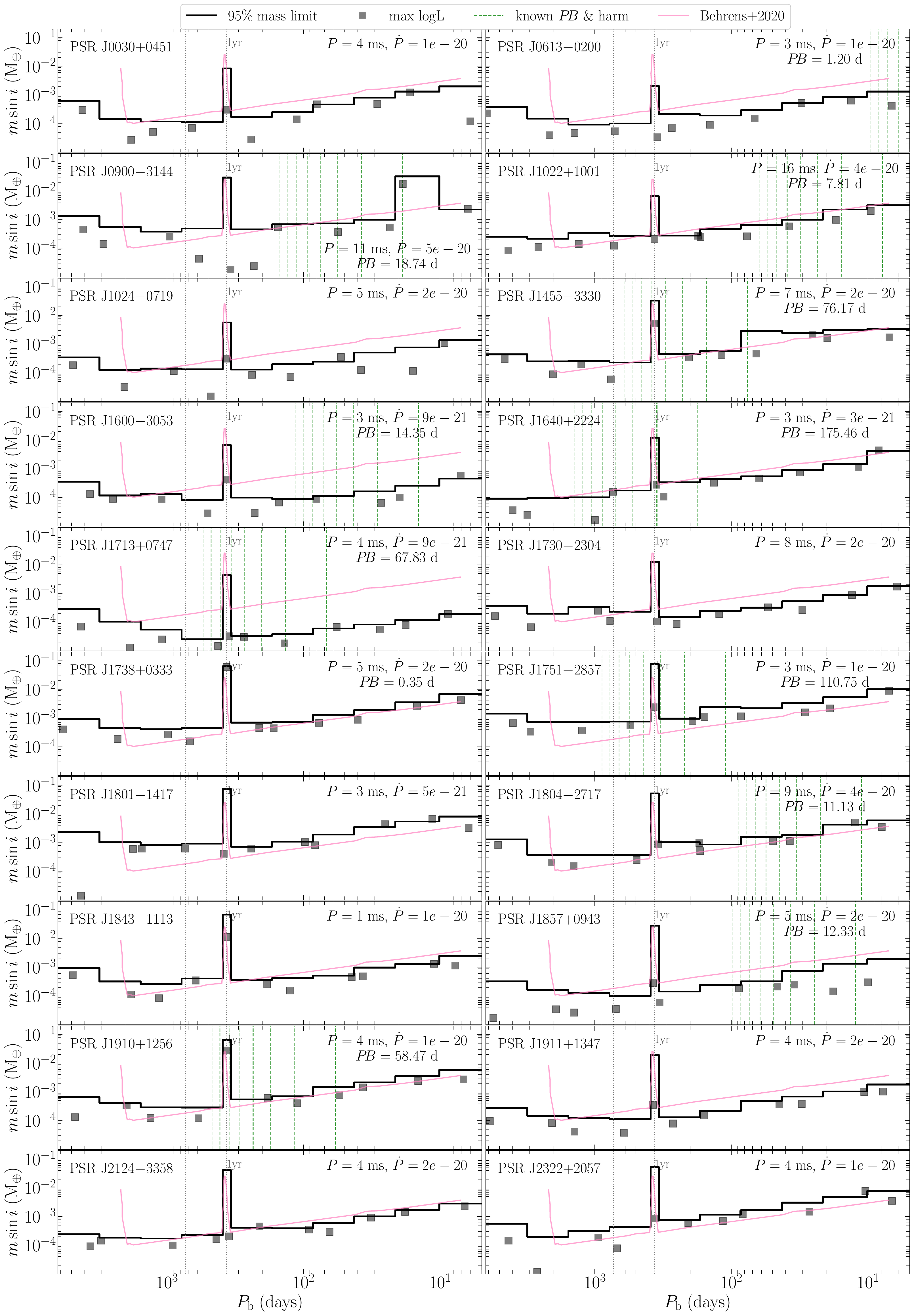}
    \caption{95\% mass limits for the 20 not-`flagged' MSPs. See the caption of Fig.~\ref{fig: flagmasslim} for more details.}
    \label{fig: notflagmasslim}
\end{figure*}

Fig.~\ref{fig: flagmasslim} shows the estimated 95\% mass limits for the 4 `flagged' MSPs, while Fig.~\ref{fig: J1909masslim} shows the mass limits of the most sensitive pulsar in this dataset, PSR J1909$-$3744. Fig.~\ref{fig: notflagmasslim} shows the results for the other 20 MSPs. Overall, the mass limits shown here are remarkably low. For example, any possible planet companion around any of the 25 MSPs is highly unlikely to have a projected mass higher than $10^{-3}\,\mathrm{M}_{\oplus}$, for orbital periods between roughly 20\,d--17\,yr. This is excluding the small area of parameter space where $P_\mathrm{b}\approx 1\,\mathrm{yr}$, where sensitivity is lost due to fitting for the proper motion and distance to the pulsar; we further ignore this in the following discussion.

PSR J1744$-$1134 -- which is a solitary MSP --- was also part of the planet search on JBO data in \citet{Nitu2022}. We can therefore directly compare the sensitivity of the JBO data by itself and the EPTA DR2, and this is shown in Fig.~\ref{fig: flagmasslim} for this pulsar; note that the JBO data are included in the EPTA DR2. As can be seen, for PSR J1744$-$1134 the EPTA timing data are more than an order of magnitude more sensitive to a planet influence than the JBO data by itself, while showing similar trends with different orbital periods.

The best mass limits found in our analysis are those estimated from the data of PSR J1909$-$3744 (shown in Fig.~\ref{fig: J1909masslim}), which constrain the 95-percentile of a planet mass to be lower than $2 \times 10^{-4}\,\mathrm{M}_{\oplus}$ for all orbital periods investigated (5\,d--17\,yr). For context, this value is approximately equal to the mass of the dwarf planet Ceres \citep[$1.6 \times 10^{-4}\,\mathrm{M}_{\oplus}$;][]{Park2019}, and roughly a tenth of the mass of Pluto \citep[$2.2 \times 10^{-3}\,\mathrm{M}_{\oplus}$;][]{Stern2015}. Furthermore, this is roughly an order of magnitude better than previously published state-of-the-art sensitivity limits, also plotted in Fig.~\ref{fig: flagmasslim} for comparison. This sensitivity curve was derived by \citet{Behrens2020} using simulations, for a `typical' MSP in the NANOGrav 11-yr dataset. Roughly half of the 25 MSPs show mass limits better than those computed by \citet{Behrens2020}, while the rest are still within one order of magnitude of these.

We note that, although we have not specifically searched for a population of `diamond' planets --- since they would be at orbital periods of $<\! 1\,\mathrm{d}$ \citep{Spiewak2018}, and this is not robust with our observational cadence --- this can still be ruled out in our dataset. This is because, as the residuals have noise of about $10\,\upmu\mathrm{s}$, any planet-like influence corresponding to an orbital period $<\!1\,\mathrm{d}$ would be obvious above the noise for a projected mass $>\!0.4\,\mathrm{M}_{\oplus}$. A diamond-planet is expected to be above a Jupiter-mass (i.e. roughly $300\,\mathrm{M}_\oplus$), which would therefore make it immediately obvious in these residuals.

\section{Fitting for a quasi-periodic Gaussian process} \label{sec: QPfit}

\subsection{Setup}
\color{black}

{It is well established that QP timing noise is prevalent in the `normal' pulsar population \citep{Hobbs2010}, thought to be due to multi-modal switching of magnetospheric processes \citep{Lyne2010}. Whilst power-law models can perform well in such conditions, strong QP noise can lead to errors in parameter estimation, and over-estimation of the power-law noise model \citep{Keith2023}.
If timing noise in MSPs is governed by similar QP processes then there is potential to better characterise the noise and consequently increase sensitivity to GW signals at the lowest frequencies.
Therefore we search for a QP process in the timing data of each of the 25 MSPs in the EPTA DR2 dataset.
The same `core' timing model is used, supplemented with the Fourier-domain Gaussian-process QP model described in \citet{Keith2023}, and implemented in \textsc{run\_enterprise}, which is outlined in the following.}

The full timing model is fit simultaneously using a Bayesian method and the MCMC sampler \textsc{emcee}, as before. The shape of the QP model used was motivated by the observed frequency-domain behaviours of slow pulsars \citep[see e.g][]{Lyne2010}. The QP effect is represented in Fourier-domain using the same infrastructure as the power-law red noise models. The characteristic PSD of the spin-frequency derivative of the pulsar, $\dot{\nu}$, is described by a sum of $N_\mathrm{harm}$ Gaussian-function terms. Each Gaussian function is centred at harmonically related frequencies $f_k = k f_\mathrm{qp}$, with $k \in \{1, \hdots, N_\mathrm{harm}\}$, where $f_\mathrm{qp} \equiv f_1$ is the fundamental frequency of the quasi-periodicity. The Gaussian functions have increasing widths, characterised by a standard deviation $\sigma_k = f_k \,\sigma = k f_\mathrm{qp} \sigma$, where $\sigma$ is a dimensionless `fractional' standard deviation, and is the same for all harmonically related terms. Furthermore, the amplitude of each Gaussian function is also decreasing exponentially, by a factor of $\exp{[-(k-1)/\lambda]}$, where in practice the value of $\lambda$ quantifies the number of significant harmonics.

Consequently, the PSD of the residuals for the QP process is described as
\begin{equation}
    P_\mathrm{\!qp}(f) = 
    \begin{cases} 
      R_\mathrm{qp}P_\mathrm{\!pl}(f_\mathrm{qp}) \, q(f)\left(\dfrac{f}{f_\mathrm{qp}}\right)^{-4} &, f\geq f_\mathrm{cut} \\
      0 &, f<f_\mathrm{cut}. 
   \end{cases}
    \label{eq: ppl_qp_fcut}
\end{equation}
for a Fourier frequency $f$ and a threshold 
\begin{equation}
f_\mathrm{cut} = 0.5 f_\mathrm{qp} \left(1-\sqrt{1-16\sigma^2}\right)   
\end{equation}
defined by the local minima of $P_\mathrm{\!qp}(f)$ to avoid an unwanted increase at very low frequencies. Further,
\begin{equation}
    P_\mathrm{\!pl}(f) = \dfrac{A_\mathrm{red}^2}{12\pi^2}\left(\dfrac{f}{1\,\mathrm{yr}^{-1}}\right)^{-\gamma_\mathrm{red}} \,\mathrm{yr}^3
    \label{eq: Ppl}
\end{equation}
is the power-law PSD for achromatic red noise (equivalent to Eq.~\ref{eq: PL}), and the quantity
\begin{equation}
    q(f) = \sum_{k=1}^{N_\mathrm{harm}} \frac{1}{k} \, \exp{\Bigg[\dfrac{-(k-1)}{\lambda}\Bigg]} \exp{\Bigg[\frac{-(f-kf_\mathrm{qp})^2}{2k^2f_\mathrm{qp}^2\sigma^2} \Bigg]}
\end{equation}
encapsulates the QP-type variability.
In practice, we set $N_\mathrm{harm}=10$ for simplicity, since we expect fewer than 10 harmonics to contribute significantly to the total signal in all cases. The $f^{-4}$ dependence in Eq.~\ref{eq: ppl_qp_fcut} is due to the transformation between the $\dot{\nu}$ and the residuals, as $\dot{\nu} \propto -\ddot{r} \propto f^{2} r$. For more details on this choice of QP model, see \citet{Keith2023}. The hyperparameters describing this process are $R_\mathrm{qp}$, $f_\mathrm{qp}$, $\lambda$, and $\sigma$. {We set uniform priors on the log of the ratio, $\log_{10}\!R_\mathrm{qp}$, the central periodicity, $1/f_\mathrm{qp}$, and on $\lambda$ and $\sigma$ as defined in Table \ref{qp_priors}. The upper bound on $1/f_\mathrm{qp}$ is chosen to prevent it becoming degenerate with the power-law noise process, and the lower bound is chosen to keep $f_\mathrm{qp}$ within the bounds of the Fourier basis.}

\begin{table}
    \centering
        \caption{Uniform prior bounds for the QP process model. $N_\mathrm{c}$ is the number of Fourier coefficients in the achromatic power-law noise model.}
    \label{qp_priors}
    \begin{tabular}{ccc}
    Parameter & Lower & Upper \\
    \hline
        $\log_{10}\!R_\mathrm{qp}$ &  $-2$ & $3.5$ \\
        $1/f_\mathrm{qp}$ & $T_\mathrm{span}/4$ & $1.2T_\mathrm{span}/N_\mathrm{c}$ \\
        $\lambda$ & $0.01$ & $10$ \\
        $\sigma$ & $10^{-3}$ & $0.2$\\
    \end{tabular}

\end{table}

\begin{figure}
    \centering
    \includegraphics[width=\linewidth]{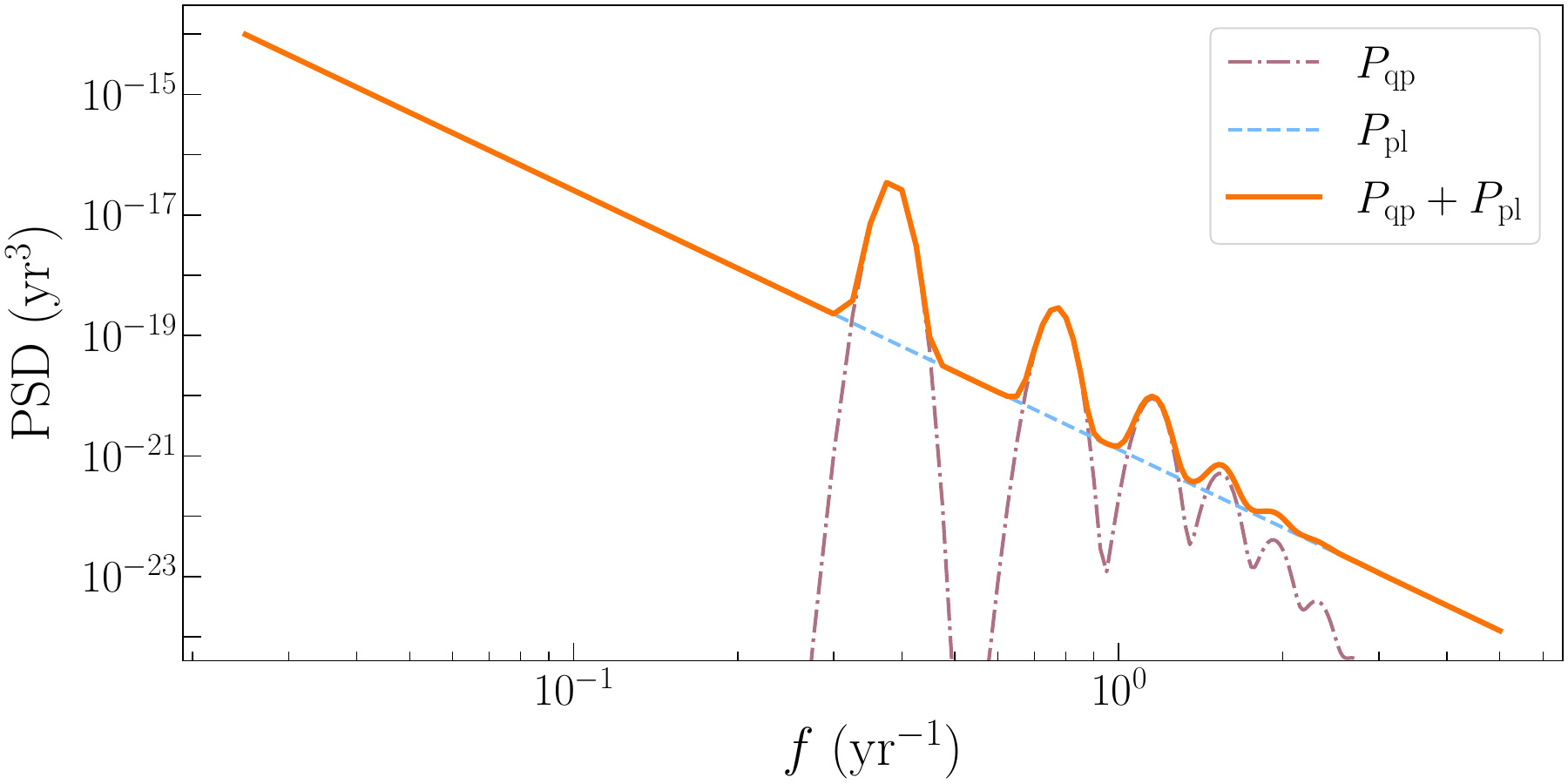}
    \caption{The functional form of the total QP and power-law PSD of the residuals, as well as its two separate components, as given by Eqs.~\ref{eq: Ppl} and~\ref{eq: ppl_qp_fcut}. The parameters used are ${A_\mathrm{pl} = 3.9 \times 10^{-10}}$, ${\gamma = 4.3}$, ${R_\mathrm{qp} = 501.2}$, ${f_\mathrm{qp} = 0.39\,\mathrm{yr}^{-1}}$, ${\sigma = 0.047}$, and ${\lambda = 0.7}$.}
    \label{fig: Ppl_and_Pqp}
\end{figure}

The QP model is designed to be used alongside the power-law red noise model, as described in Section~\ref{sec: coretiming}, such that the full achromatic red noise is modelled according to a PSD of $P_\mathrm{\!qp} + P_\mathrm{\!pl}$. Fig.~\ref{fig: Ppl_and_Pqp} illustrates the shape of the functional form of $P_\mathrm{\!qp}$ and $P_\mathrm{\!pl}$ for some example parameters.

\color{black}
\subsection{Results \& Discussion}
\begin{figure} 
    \centering
    \includegraphics[width=0.49\linewidth]{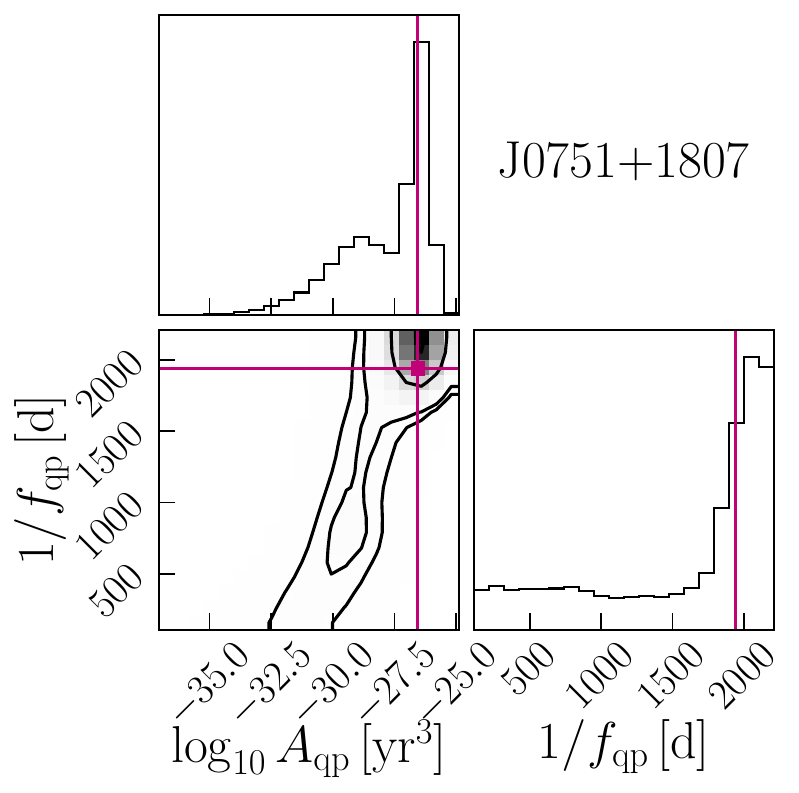}
    \includegraphics[width=0.49\linewidth]{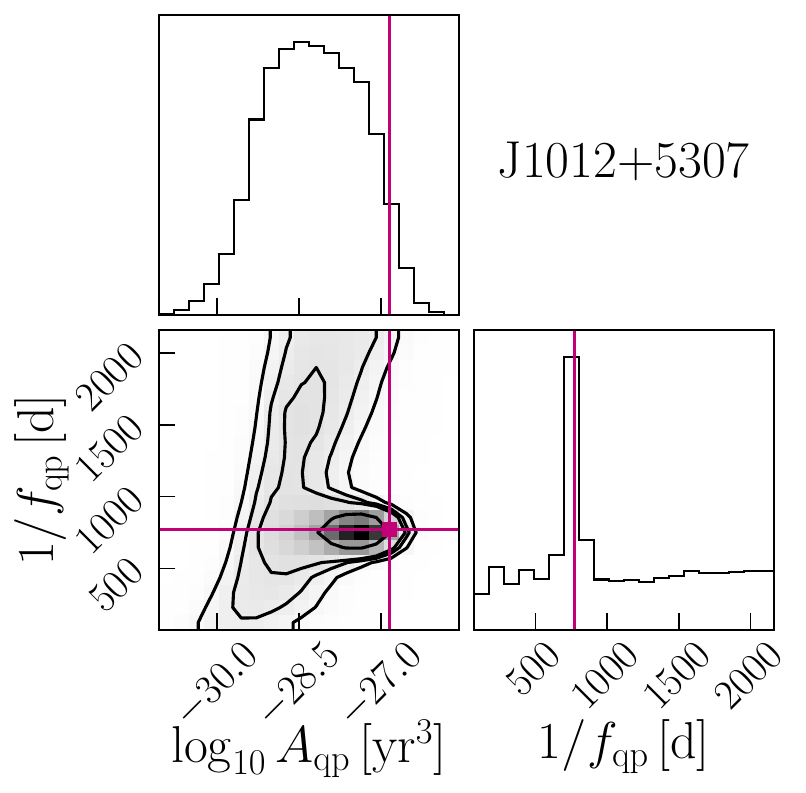} \\
    \includegraphics[width=0.49\linewidth]{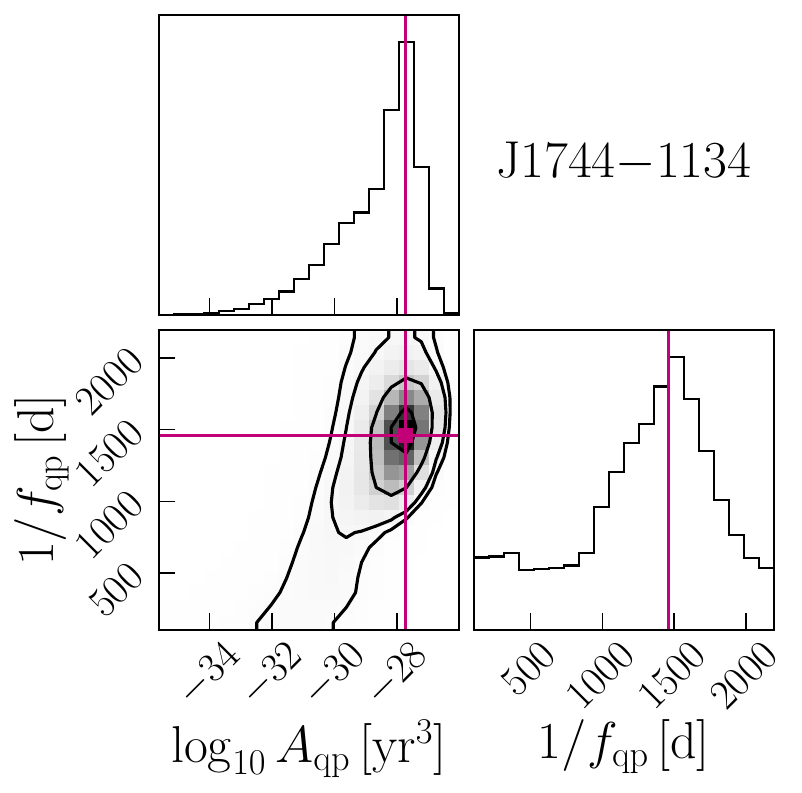}
    \includegraphics[width=0.49\linewidth]{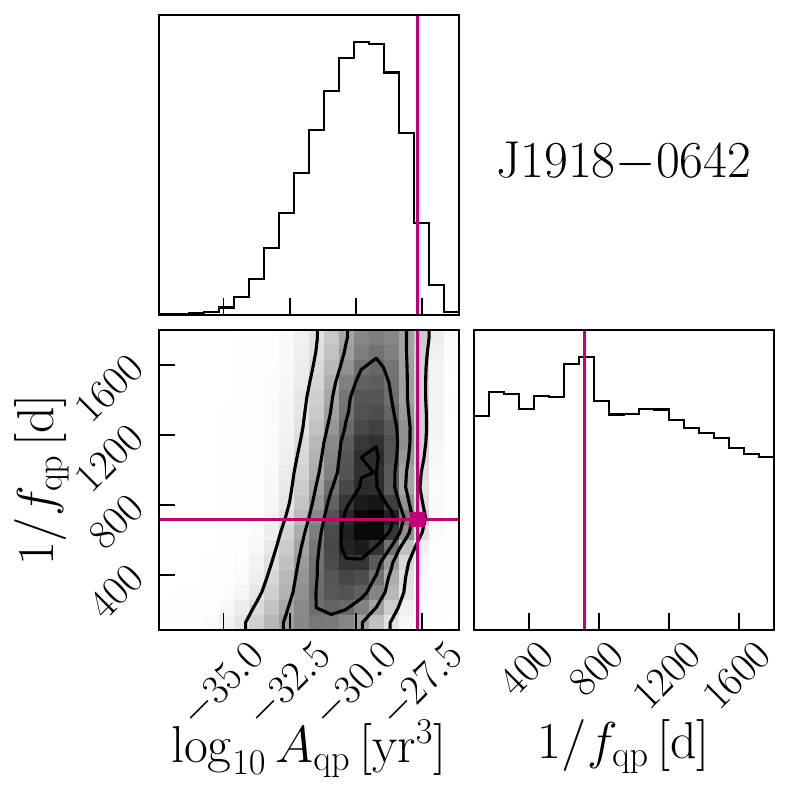}
    \caption{A subset of the corner-plots for the 4 pulsars highlighted in the periodicity search of Section~\ref{sec: planetfits}. The properties of the QP model fitted are summarised through the fundamental periodicity $1/f_\mathrm{qp}$ and the power amplitude at this corresponding Fourier frequency. The contour levels are at the (0.5, 1, 1.5, 2)-sigma equivalent, as used in the library \textsc{corner.py}. The purple vertical lines correspond to the maximum-likelihood values of each parameter.}
    \label{fig: QP_cornerplots}
\end{figure}
In short, our analysis concluded that none of the MSPs are well characterised by our QP model. In all cases, the posterior distributions of the parameters $\lambda$ and $\sigma$ are almost identical to their respective uniform prior. A similar situation was seen for parameter $R_\mathrm{qp}$, which quantifies the strength of the QP power compared to the power-law red noise power. While the prior for $R_\mathrm{qp}$ was chosen to be log-uniform, its posterior distribution generally recovers this prior, but with a decrease at large values; the only exception to this is perhaps the case of PSRs J0751$+$1807 and J1744$-$1134. This is not unexpected, as it illustrates a (not very constraining) upper limit for any existing QP-type power; the posterior distributions suggest that it is unlikely that $R_\mathrm{qp} > 10^3$ for any of the studied MSPs.

The fourth hyperparameter, $f_\mathrm{qp}$, describes the fundamental frequency of the QP process. For most MSPs in this analysis, the posterior of $f_\mathrm{qp}$ is also unconstraining. However, for PSRs J0751$+$1807, J1012$+$5307, and J1744$-$1134 there was some level of preference for particular frequencies. These pulsars are included in the subset of 4 MSPs that were `flagged' in the planet search in Section~\ref{sec: 25PSRs_flagged}. We therefore concentrate on the results of the QP fitting for the previously considered 4 MSPs. The most interesting, and most constrained properties of the QP model are the fundamental period of the QP process (given by $1/f_\mathrm{qp}$), and the power amplitude at this Fourier frequency. According to Eqs.~\ref{eq: ppl_qp_fcut} and~\ref{eq: Ppl}, we can write this amplitude at $f_\mathrm{qp}$ as
\begin{equation}
    A_\mathrm{qp} = R_\mathrm{qp} \dfrac{A_\mathrm{red}^2}{12\pi^2} \left(\dfrac{f_\mathrm{qp}}{1\,\mathrm{yr}^{-1}}\right)^{-\gamma_\mathrm{red}},
\end{equation}
Thus we straight-forwardly derive the posterior of $\log_{10}\!A_\mathrm{qp}$ from the posteriors of $\log_{10}\!R_\mathrm{qp}$, $\log_{10}\!A_\mathrm{red}$, $f_\mathrm{qp}$, and $\gamma_\mathrm{red}$. {We note that although the prior on $\log_{10}\!R_\mathrm{qp}$ is uniform, the effective prior on the derived parameter $\log_{10}\!A_\mathrm{qp}$ is more complex.} Fig.~\ref{fig: QP_cornerplots} shows the posteriors of $1/f_\mathrm{qp}$ and $\log_{10}\!A_\mathrm{qp}$ and the relationship between them for the 4 mentioned pulsars. Of the 4 MSPs, PSR J1012$+$5307 shows the clearest peak in the $1/f_\mathrm{qp}$ posterior. The corresponding posterior of PSR J0751$+$1807 shows a preference for the maximum of the prior, which is set to be a quarter of the total observing span. Further, the QP-period posterior of J1744$-$1134 is wide, but nearly Gaussian-shaped. On the other hand, the posterior of PSR J1918$-$0642 does not significantly differ from the prior. None of the amplitude posteriors of the 4 MSPs are strictly Gaussian shaped, although the results for PSRs J1012$+$5307 and J1918$-$0642 are reasonably close. For PSRs J0751$+$1807 and J1744$-$1134, the posterior distributions have a long tail at low amplitudes, with an abrupt cut-off near $A_\mathrm{qp} \sim 10^{-26} \mathrm{yr}^3$.

Furthermore, Table~\ref{tab: flaggedPSRs} shows a direct comparison between the period found in the planet fitting, and the inverse of the maximum-likelihood fundamental frequency of this QP fitting. The standard deviation estimated from the posterior of $1/f_\mathrm{qp}$ is also included to give an idea of the width of the posterior distribution, but this should be considered in conjunction with the shape of the posterior, which is generally not exactly Gaussian (as shown in Fig.~\ref{fig: QP_cornerplots}). From the results in Table~\ref{tab: flaggedPSRs}, we can see that in all but PSR J1918$-$0642 the maximum-likelihood QP fundamental periodicity is similar to the planetary orbital period. 
In the case of PSRs J0751$+$1807 and J1744$-$1134, the large periodicity as a fraction of the total data span suggests that both the planet and the QP fitting are trying to account for the large power observed at high timescales. For PSRs J1012$+$5307, the periodicity of approximately 2\,yr is found in the QP fitting as well, although not well described by this QP model. 

\section{Conclusions} \label{sec: conclusions}

Overall, we conclude that none of the timing models of the 25 MSPs in the EPTA DR2 would clearly benefit from the addition of a QP/periodic process. In particular, the lack of a detectable Fourier-domain QP Gaussian process in these data means we do not expect there to be a bias in parameters or noise estimates due to this kind of behaviour, as was found by \citet{Keith2023}.

{Four pulsars are highlighted as showing potential periodic or QP processes in our analysis.
Two (PSRs J0751$+$1807 and J1918$-$0642) are pulsars for which the original EPTA analysis selected against achromatic red noise, but we find it likely that the periodic signal we measure is caused by unmodelled red noise.
The other two (PSRs J1744$-$1134 and J1012$+$5307) are among the pulsars suggested to have `complex' behaviours in the EPTA noise analysis \citep{EPTA2023II}.
Our analysis also hints that a pure power-law process may not be sufficient, however neither the periodic nor the QP models we trialled seem to meaningfully improve the results.
These pulsars will certainly continue to be studied in the near future with EPTA and IPTA analyses, as we attempt to better specify the noise models for PTAs.}

Finally, the planet-fitting analysis on these 25 MSPs allowed us to put highly constraining limits on the masses of any planetary companions orbiting these pulsars. The timing data of PSR J1909$-$3744, which yielded the best mass limits, allowed us to constrain the 95-percentile to approximately the mass of the dwarf planet Ceres ($\sim\!\!2\times 10^{-4}\,\mathrm{M}_{\oplus}$) for orbital periods between 5\,d--17\,yr. These limits are more than an order of magnitude improved compared to the previous sensitivity curve estimated by \citet{Behrens2020} for the NANOGrav 11-yr data, and are the best planet-mass limits from pulsar timing to date.

\section*{Acknowledgements}
% Pulsar research at Jodrell Bank is supported by a consolidated grant from the UK Science and Technology Facilities Council (STFC). 
ICN was supported by the STFC doctoral training grant ST/T506291/1. 

Part of the EPTA data used in this work is based on observations with the 100-m telescope of the Max-Planck-Institut f\"{u}r Radioastronomie (MPIfR) at Effelsberg in Germany. Pulsar research at the Jodrell Bank Centre for Astrophysics and the observations using the Lovell Telescope are supported by a Consolidated Grant (ST/T000414/1) from the UK's Science and Technology Facilities Council (STFC). The Nan{\c c}ay radio Observatory is operated by the Paris Observatory, associated with the French Centre National de la Recherche Scientifique (CNRS), and partially supported by the Region Centre in France. We acknowledge financial support from ``Programme National de Cosmologie and Galaxies'' (PNCG), and ``Programme National Hautes Energies'' (PNHE) funded by CNRS/INSU-IN2P3-INP, CEA and CNES, France. We acknowledge financial support from Agence Nationale de la Recherche (ANR-18-CE31-0015), France. The Westerbork Synthesis Radio Telescope is operated by the Netherlands Institute for Radio Astronomy (ASTRON) with support from the Netherlands Foundation for Scientific Research (NWO). The Sardinia Radio Telescope (SRT) is funded by the Department of University and Research (MIUR), the Italian Space Agency (ASI), and the Autonomous Region of Sardinia (RAS) and is operated as a National Facility by the National Institute for Astrophysics (INAF). 

%%%%%%%%%%%%%%%%%%%%%%%%%%%%%%%%%%%%%%%%%%%%%%%%%%
\section*{Data Availability}
The EPTA DR2 data underlying the work in this paper are available at \href{https://doi.org/10.5281/zenodo.8164424}{https://doi.org/10.5281/zenodo.8164424}, as per \citet{EPTA2023I}.

%The inclusion of a Data Availability Statement is a requirement for articles published in MNRAS. Data Availability Statements provide a standardised format for readers to understand the availability of data underlying the research results described in the article. The statement may refer to original data generated in the course of the study or to third-party data analysed in the article. The statement should describe and provide means of access, where possible, by linking to the data or providing the required accession numbers for the relevant databases or DOIs.

%%%%%%%%%%%%%%%%%%%% refERENCES %%%%%%%%%%%%%%%%%%


\bibliographystyle{mnras}
\bibliography{bibliography} % if your bibtex file is called example.bib

%\begin{thebibliography}{99}
%\end{thebibliography}

%%%%%%%%%%%%%%%%%%%%%%%%%%%%%%%%%%%%%%%%%%%%%%%%%%

%%%%%%%%%%%%%%%%% APPENDICES %%%%%%%%%%%%%%%%%%%%%

% \appendix
% \section{Mass limits on the non-flagged MSPs} \label{app: notflagmasslim}

%%%%%%%%%%%%%%%%%%%%%%%%%%%%%%%%%%%%%%%%%%%%%%%%%%

% Don't change these lines
\bsp	% typesetting comment
\label{lastpage}
\end{document}